\documentclass[english,preprint2,english]{aastex}
\usepackage[T1]{fontenc}
\usepackage[latin1]{inputenc}
\usepackage{graphicx}

\makeatletter
\usepackage{times}
\usepackage{babel}
\makeatother
\begin{document}

\title{Fluxon Modeling of Low-Beta Plasmas}

\affil{\emph{(In Press: Journal of Atmospheric and Solar-Terrestrial
Physics; accepted Oct 2005)}}

\author{C.E. DeForest}

\affil{Southwest Research Institute, 1050 Walnut Street Suite 400, Boulder,
CO 80302, USA}

\author{C.C. Kankelborg}

\affil{Montana State University, Bozeman, MT, USA}

\begin{abstract}
We have developed a new, quasi-Lagrangian approach for numerical modeling
of magnetohydrodynamics in low to moderate $\beta$ plasmas such as
the solar corona. We introduce the concept of a {}``fluxon'', a
discretized field line. Fluxon models represent the magnetic field
as a skeleton of such discrete field lines, and interpolate field
values from the geometry of the skeleton where needed, reversing the
usual direction of the field line transform. The fluxon skeleton forms
the grid for a collection of 1-D Eulerian models of plasma along individual
flux tubes. Fluxon models have no numerical resistivity, because they
preserve topology explicitly. Our prototype code, \emph{FLUX}, is
currently able to find 3-D nonlinear force-free field solutions with
a specified field topology, and work is ongoing to validate and extend
the code to full magnetohydrodynamics. FLUX has significant
scaling advantages over conventional models: for {}``magnetic carpet''
models, with photospheric line-tied boundary conditions, FLUX
simulations scale in complexity like a conventional 2-D grid although
the full 3-D field is represented. The code is free software and is
available online. In this current paper we introduce fluxons and our
prototype code, and describe the course of future work with the code.\\
~\\
~
\end{abstract}

\section{Introduction}

MHD modeling is key to understanding solar eruptive events and their
effect on the heliospheric enviroment. Solar flares are known to be
driven by magnetic reconnection (e.g. \citealt{Sturrock1984}) and
coronal mass ejections (CMES) are generally tied to magnetic instability
of one kind or another (e.g. \citealt{Sturrock1989,Amari1999,Antiochos1999,Chen2000,Fan2003,Roussev2003}).
Understanding and predicting such events requires numeric modeling
of the plasma in the solar corona as it evolves under the line-tied
boundary conditions imposed by the solar photosphere. Current MHD
modeling software is not able to reproduce the conditions of the solar
corona under controlled conditions, because numerical effects in the
simulation dominate over their counterparts in the real corona for
many physical situations. This is evident in the difficulty of maintaining
strong current sheets or other stressed flux systems, such as filaments,
for long periods of time; and in the difficulty of reproducing the
rapidly varying heating rate in coronal loops and bright points.

Conventional magnetohydrodynamic models typically use Eulerian (fixed)
grids, introducing nonphysical dissipative effects such as viscosity
and resistivity. These dissipative effects dominate the behavior of
quiescent structures with current sheets and may even destabilize
simulated CME-bearing systems \citet{Lin2002}, making modeling of
solar evolution difficult at best. Further, numerical diffusion of
both momentum ('numerical viscosity') and magnetic field ('numerical
resistivity') are dependent on grid speed, so that Eulerian grid size
is typically chosen to oversample the physical structure. The extra
resolution is used to minimize the unwanted diffusion and preserve
the plasma system for as much simulated time as possible. Oversampling
by a factor of 10 in each dimension yields a factor of $10^{4}$ slowdown
in the overall simulation, requiring huge facilities to simulate even
simple systems in 3-D. Adaptive-mesh refinement improves performance
significantly (e.g. \citealt{Welsch2004,Lynch2003}) but still requires
oversampling.

Lagrangian grids eliminate numerical resistivity but add additional
problems. For example, as the grid distorts with the plasma motion,
the fidelity of discrete differential operators degrades. Fully Lagrangian
treatments of the corona shear rapidly because fluid motion is decoupled
in the cross-field direction.

To maximize the advantages of both the Eulerian and Lagrangian approaches
to MHD modeling, we have developed a prototype fluxon modeling code,
\emph{FLUX} (the {}``Field Line Universal relaXer''), that is a
hybrid between the two. In the low-$\beta$ regime, all forces are
negligible when compared to the Lorenz force components; FLUX is essentially
a force-free field solver that can support an independent plasma density
parameter at each location in the simulation. FLUX demonstrates the
fluxon method, and work is ongoing to add plasma pressure and related
physical phenomena to the simulation framework. In the following sections,
we briefly describe the basis of the fluxon numerical approach (\S\ref{sec:Fluxon-theory}),
describe our code and its performance (\S\ref{sec:FLUX-implementation}),
and discuss the direction of future work both on code development
and on applications (\S\ref{sec:Future-work}).

\section{\label{sec:Fluxon-theory}Fluxon theory and implementation}

The basis of the fluxon approach to numerical modeling is the analogy
between field lines and an associated vector field: a field line map
completely describes the associated magnetic field, and vice versa.
This analogy has normally been used to visualize the magnetic field:
the field vector value is calculated everywhere on a grid of values,
and then interpolated to {}``shoot'' field lines through the grid
for visualization. But the analogy works in the reverse direction
too: every physical property of the magnetic field may be described
in terms of the behavior of individual magnetic field lines. In highly
conductive plasmas, the field line description takes on more utility
and meaning than in resistive physical systems, because field line
topology is preserved under ideal MHD. 

\begin{figure}[tb]
\includegraphics[%
  width=1.0\columnwidth,
  keepaspectratio]{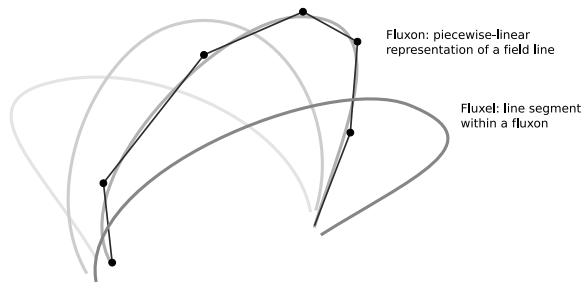}

\caption{\label{cap:fluxon}A \emph{fluxon} representation of the magnetic
field replaces smooth field lines (each of which represents an infinitesimal
amount of flux) with piecewise-linear fluxons (each of which represents
a small but finite amount of flux). Each fluxon is composed of flux
elements, or \emph{fluxels}, that represent small units of field-aligned
length $ds$.}
\end{figure}

FLUX is currently a relaxation solver for force-free magnetic systems:
the magnetic field is represented as a skeleton of piecewise linear
curves, \emph{fluxons}, each of which represents a finite quantum
of magnetic flux contained in a thin volume around a central curve.
Fluxons differ from conventional field lines in that a field line
represents an infinitesimal amount of magnetic flux, while a fluxon
represents a discrete, finite amount of magnetic flux; the word is
also used, with approximately the same meaning, in the context of
quantized magnetic systems such as Josephson junctions and quantum
computers (e.g. \citealt{Calidonna2005,Ustinov1993}). The magnetic
field is considered to be nearly parallel to the fluxon everywhere
in its neighborhood, so that each fluxon may be considered to represent
a non-twisted flux tube and (as with conventional field lines) magnetic
field strength may be calculated by determining the areal density
of fluxons that pass through a plane perpendicular to the field direction.
Other quantities such as field gradients may be calculated from the
local geometry of the fluxons.

\begin{figure}[tb]
\center{\includegraphics[%
  width=0.35\columnwidth,
  keepaspectratio]{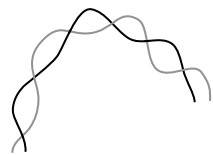}}

\caption{\label{cap:currents}Two-fluxon system demonstrating representation
of current. A computable current runs along the axis of the twisted
loop.}
\end{figure}

\begin{figure}[tb]
\center{\includegraphics[%
  width=0.40\columnwidth,
  keepaspectratio]{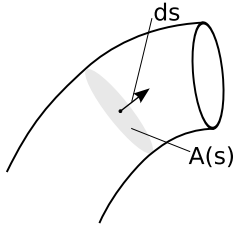}}

\caption{\label{cap:lorenz-cartoon}A curved flux tube carries flux $\Phi$
along its length; its geometry is parametrized by path length $s$
and its cross section is $A(s)$.}
\end{figure}

Each fluxon in a simulation arena is composed of an ordered collection
of \emph{flux elements}, or \emph{fluxels} (Figure \ref{cap:fluxon}).
A fluxel represents a small amount of field-aligned length $ds$,
and the differential MHD equations are discretized using $d\Phi\Rightarrow\Phi_{0}$
(where $\Phi_{0}$ is the quantum of magnetic flux) and $ds\Rightarrow\Delta s$
(when calculating curvature and similar quanties). Each fluxel takes
the same part in a fluxon simulation that a pixel (or voxel or grid
element) takes in a conventional Eulerian simulation. In the current
implementation, fluxels are line segments and their associated fluxons
are thus piecewise-linear. One may imagine spline or other curvilinear
interpolation, but the geometric calculations are greatly simplified
by a piecewise linear representation.

Computationally, fluxons are represented with dynamically allocated
data structures: each fluxon is a linked list of fluxels, each of
which contains position information for one of the two endpoints of
the fluxel, and some ancillary data used to calculate local geometry
and forces. In a relaxation calculation, the fluxel positions are
relaxed to find a force-free equilbrium. At each relaxation step,
all included forces are calculated at each fluxel node, and the node
takes a step in the direction of the vector sum of the forces acting
on it. The motions and force laws are constructed in such a way that
fluxels never cross one another, unless forced to by the physics of
the model. This ensures that magnetic topology is conserved: the discrete
nature of the fluxon skeleton eliminates numerical reconnection. Including
just the Lorenz forces in the relaxation yields approximations to
non-linear force free field solutions with prescribed topology.

A single fluxon cannot by itself represent a field-aligned current;
but multiple fluxons can represent currents through twist (Figure
\ref{cap:currents}). The differential quantity $\vec{\nabla}\times\vec{B}$
may be estimated directly from the discrete geometry of nearby field
lines.

In the remainder of this section, we describe the theory and implementation
used to find the Lorenz force throughout the simulation and hence
to relax toward a force-free solution.

\subsection{\label{sub:A-geometric-approach}A geometric approach to the Lorenz
force}

Calculating force free fields requires formulating the Lorenz magnetic
force in terms of the geometry of the fluxon grid, which represents
the magnetic field as a collection of small magnetic flux tubes. We
here derive the familiar force law $\overrightarrow{F_{L}}=\overrightarrow{J}\times\overrightarrow{B}$
from the energetics of a small, discrete flux tube, to demonstrate
that the familiar forces acting on infinitesimal field lines can be
represented using only the geometry of finite flux tubes. Consider
the magnetic energy $E_{B}$ of a finite, curved flux tube that carries
a magnetic flux $\Phi$ and whose shape and cross section $A$ are
parametrized by path length $s$ along the tube, as in Figure \ref{cap:lorenz-cartoon}.
(Note that the cross-section may be an arbitrary shape, not only round
as depicted here). If the magnetic field is constant across the cross-section
of the tube, then the magnetic energy $E_{B}$ is given by\begin{equation}
E_{B}=\int\frac{{B^{2}}}{8\pi}d^{3}V=\frac{\Phi^{2}}{8\pi}\int\frac{ds}{A(s)}\label{eq:energetics}\end{equation}
where one factor of $A$ is cancelled from the denominator by carrying
out the cross-sectional part of the volume integral. Taking the differential
along the length of the flux tube lets us characterize the energy
per unit length:\begin{equation}
dE_{B}=\frac{\Phi^{2}}{8\pi A(s)}ds=\frac{\Phi B(s)}{8\pi}ds.\label{eq:differential}\end{equation}
Differentiating with respect to displacement $x_{i}$ of a point on
the flux tube yields the differential of the $i^{th}$component of
the Lorenz force. \begin{equation}
dF_{i}=-\frac{\partial\left(dE_{B}\right)}{\partial x_{i}}=-\frac{\Phi}{8\pi}\left(\frac{\partial(ds)}{\partial x_{i}}B+\frac{\partial B}{\partial x_{i}}ds\right),\label{eq:differential-force}\end{equation}
where the first term is due to the variation of the path length from
the displacement and the second term is due to the gradient of $B$.
Noting that the path length variation is just the dot product of the
displacement with $ds$ times the negative curvature $-(d/ds)(\partial s/\partial x_{i})$
allows us to interchange the differentials, at the expense of a sign
change.

\begin{equation}
\frac{dF_{i}}{ds}=-\frac{\Phi}{8\pi}\left(-B\frac{d}{ds}\frac{\partial s}{\partial x_{i}}+\frac{dB}{dx_{i}}\right)\label{eq:force}\end{equation}
Breaking the total derivative $d/ds$ into partial derivative terms
gives:\begin{equation}
\frac{dF_{i}}{ds}=\frac{\Phi}{8\pi}\left(\left(\sum B\frac{\partial x_{i}}{\partial s}\frac{\partial}{\partial x_{i}}\right)\left(\frac{\partial s}{\partial x_{i}}\right)-\frac{dB}{dx_{i}}\right)\label{eq:partials}\end{equation}
 which reproduces the familiar curvature and pressure force terms.
Separating out $\Phi$ into $BA$ and converting to vector notation
gives\begin{equation}
\frac{d\vec{F}}{ds}=\frac{A}{8\pi}\left(\left(\vec{B}\cdot\vec{\bigtriangledown}\right)\vec{B}-\frac{\vec{\nabla}B^{2}}{2}\right)\label{eq:Lorenz}\end{equation}
 where we have taken advantage of the relation $\vec{\bigtriangledown}\cdot\vec{B}=0$
to commute the scalar $B$ through the $\vec{B}\cdot\vec{\bigtriangledown}$
operator. Equation \ref{eq:Lorenz} is just the familiar Lorenz force
relation, multiplied by the cross section of the flux tube. The left-hand
term is the {}``curvature force'' and the right-hand term is {}``pressure
force''. Equation \ref{eq:Lorenz} may be more cleanly expressed
by collecting terms:\begin{equation}
\frac{8\pi}{\Phi}\frac{d\vec{F}}{ds}=\left(\hat{B}\cdot\vec{\bigtriangledown}\right)\vec{B}-\vec{\bigtriangledown}B\label{eq:Lorenz-clean}\end{equation}
 Finally, for force-free calculations it is useful to divide out the
magnitude $B$ of the magnetic field, yielding the field-normalized
Lorenz force:

\begin{equation}
\frac{8\pi}{AB^{2}}\frac{d\vec{F}}{ds}=\left(\hat{B}\cdot\vec{\bigtriangledown}\right)\hat{B}-\frac{\vec{\bigtriangledown}B}{B}\label{eq:normalized-Lorenz}\end{equation}
 Setting the left-hand side of either Equation \ref{eq:Lorenz-clean}
or Equation \ref{eq:normalized-Lorenz} to zero describes a force-free
magnetic field, but Equation \ref{eq:normalized-Lorenz} is especially
useful because the curvature force is represented entirely in terms
of the local curvature of the flux tube without reference to the field
strength $B$, and the magnetic pressure force is also reducible to
simple form. We refer to the left hand term as $\mathcal{F}_{cn}$,
the field-normalized curvature force per unit length; and to the right
hand term as $\mathcal{F}_{pn}$, the field-normalized pressure force
per unit length.

\subsection{\label{sub:Discretizing-the-Lorenz}Discretizing the Lorenz force
with fluxons}

The differential quantities in \S \ref{sub:A-geometric-approach}
must be discretized for use with fluxons: each fluxon is a set of
piecewise linear curves. Each fluxel has a finite length $l$ rather
than a differential length $ds$, with vertices at each end. What
is desired is not the force per unit length along each fluxon, but
the force acting on each vertex. The curvature and pressure force
are discretized slightly differently because the curvature of a fluxon
is defined only at the vertices, while the pressure is only defined
near the center of each line segment. 

The normalized curvature force is simple to calculate, because of
a fortunate cancellation: the amount of curvature from fluxel center
to fluxel center is inversely related to the lengths of the fluxels,
canceling the length factor in the integral. The B-normalized curvature
force at each vertex $v$ is thus proportional to the offset angle
at the vertex between two fluxels, as can be seen by integrating $\left(\hat{B}\cdot\vec{\bigtriangledown}\right)\hat{B}$
along the line segments between $v$ and its neighbors $v-1$ and
$v+1$: $ $\begin{equation}
F_{cn,v}=\frac{1}{2}\int_{v-1}^{v+1}\mathcal{F}_{cn}ds=\left(\frac{l}{2}\right)\left(\frac{2\Delta\theta_{v}}{l}\right)=\Delta\theta_{v}\label{eq:curvature-force}\end{equation}
where $l$ is the total line segment length and $\Delta\theta_{v}$
is the total amount of bend at vertex $v$.

The other half of the Lorenz force, the field-normalized pressure
force, requires characterizing the geometry of each fluxon's neighborhood
to determine the magnetic pressure gradient. While the complete magnetic
pressure $\vec{\bigtriangledown}B$ is of interest in full MHD simulations,
we note that displacement of a field line (and hence fluxon) along
the direction of the field makes no physical change in the absence
of plasma forces, and hence we use only the perpendicular pressure
gradient $\vec{\bigtriangledown}_{\perp}B$ to find the force-free
equilibrium.

The flux associated with each fluxon $F$ is considered to occupy
the locus that is closer to $F$ than to any other. This locus is
the \emph{Voronoi neighborhood} or \emph{Voronoi cell} of the fluxon,
and a collection of fluxons, considered as curvilinear manifolds in
three dimensions, forms a \emph{Voronoi foam} of such cells. There
is a large body of literature on computerized \emph{Voronoi analysis,}
the process of calculating and characterizing such neighborhoods near
various families of manifolds in two, three, and more dimensions;
interested readers are referred to \citet{PreparataShamos1985} for
a good introduction to the subject. 

The geometry of the Voronoi cell of the fluxon completely describes
the variation in the field strength B of the flux tube represented
by the fluxon: the cross-sectional area of the cell gives $B$, and
the asymmetry determines $\vec{\bigtriangledown}_{\perp}B$, near
the fluxon. 

\begin{figure}[tbh]
\begin{center}\includegraphics[%
  height=2.5in,
  keepaspectratio]{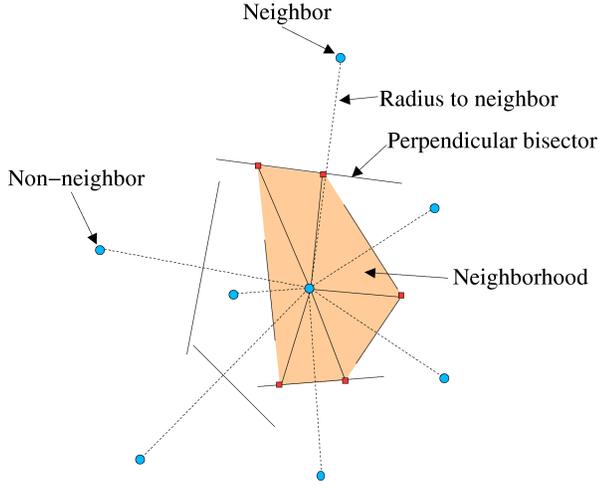}\end{center}

\caption{\label{cap:voronoi-construction}Construction of a 2-D Voronoi cell.
See text for discussion.}
\end{figure}

The Voronoi neighborhood of a one-dimensional piecewise linear manifold
(such as a fluxon) embedded in three dimensions is described by a
family of spliced fourth-order bivariate polynomials (see, e.g., \citealt{PreparataShamos1985}
and references therein); solving such curves is computationally expensive.
We instead approximate the Voronoi neighborhood of each fluxel along
the fluxon with a prism extruded along the length of the fluxel, and
determine the two-dimensional Voronoi neighborhood in the cross section
of the prism. When projected into the cross-sectional plane of a central
fluxel, the nearby fluxels appear as points, and the construction
is straightforward (as illustrated in Figure \ref{cap:voronoi-construction}).
A line segment is constructed from the central fluxel to each nearby
fluxel, and the perpendicular bisector of each such segment is found.
The smallest convex polygon that can be constructed from the bisectors
and that also includes the central fluxel is the two-dimensional Voronoi
cell of the central fluxel. The fluxels whose bisectors contribute
to the shape of the Voronoi cell are the \emph{neighbors} of the central
fluxel, and a list of them is retained. After an initial seeding step,
the Voronoi calculation at each relaxation step involves only the
neighbors and next-nearest-neighbors from the previous step, speeding
the calculation. In any sufficiently large field of points, the average
number of edges in each Voronoi cell (and hence neighbors of the central
fluxel) converges to 6, so the Voronoi calculation for each cell runs
in constant average time and for the entire simulation it runs in
$O(n)$ time.

The projection function we use to project each fluxel into the cross-sectional
plane is not Cartesian: we find the point on the candidate fluxel
that is closest to the central fluxel, project that point into two
dimensions, and then multiply the radial distance by the fourth power
of the secant of the out-of-plane angle to the candidate, artificially
raising the distance to fluxels that are out of the cross-sectional
plane. This is a smooth way of selecting the fluxels of most interest
-- those near the perpendicular plane of the central fluxel. Fluxels
that are far out of the plane are projected at a farther distance,
so that they are usually not close enough to become neighbors during
the cell construction. Fluxels are not permitted to interact with
the previous and next fluxel on the same fluxon, because the distance
to those fluxels is ill-defined (0/0 discontinuity). 

The out-of-plane radial scaling function does not affect relaxation
results strongly, provided that it is symmetric and grows fast enough:
once a nearby projected fluxel is removed far enough from the origin,
it is no longer considered a neighbor and does not affect the local
force calculation. The fluxels of interest are those near the perpendicular
plane of the central fluxel, and there the scaling function is near
unity.

Once the Voronoi geometry is known, the magnetic field still remains
to be calculated. We are free to choose any non-pathological distribution
of flux within the Voronoi cell, as the cells are by definition smaller
than the physical resolution of the model. For analytic convenience,
we treat the magnetic field in the Voronoi cell as being in \emph{sectorwise
angular equipartition}: the cells are described as a collection of
triangular segments, each of which has uniform magnetic field and
each of which has a total amount of flux that is proportional to the
angle subtended by the segment. This prevents currents from forming
at the boundaries between parallel fluxons (by construction, the field
strength is equal on opposite sides of such a boundary), allows current
sheets to form between nonparallel fluxons, and yields simple formulas
for the average field and the perpendicular field gradient. This assumption
yields a simple formula for the field-normalized magnetic pressure
force $F_{pn}$:\begin{equation}
\vec{F}_{pn}=\int_{v}^{v+1}\mathcal{F}_{pn}ds=l\frac{\left\langle \bigtriangledown_{\perp}B\right\rangle }{\left\langle B\right\rangle }=l\sum_{i}\frac{\hat{n_{i}}\Delta\phi_{i}}{\pi r_{i}}\label{eq:gradient}\end{equation}
where $\hat{n_{i}}$is the direction to the $i^{th}$ neighbor, $\Delta\phi_{i}$
is the angle subtended by the corresponding edge of the Voronoi cell,
$r_{i}$ is the distance of closest approach of the corresponding
perpendicular bisector, and $l$ is the length of the line segment
from $v$ to $v+1$. The formula works even in the case of open Voronoi
cells (which are not closed polygons), because although the field
is considered to be identically zero in the open directions, it is
nonzero in the closed directions. This prevents the magnetic pressure
force from being identically zero on the outermost field lines of
the simulation (which usually have open Voronoi cells).

\subsection{Grid relaxation to find equilibrium solutions}

At each relaxation step, FLUX calculates the field-normalized curvature
force at each node, and the pressure force at the center of each fluxel.
The pressure forces for the leading and trailing fluxels for each
node are averaged together, to produce an average pressure force at
the node. The average pressure force, in turn, is added to the curvature
force at the node to find a normalized total force. At each relaxation
step, all nodes are moved in the direction of the corresponding normalized
total force, until a relaxation condition is met. This type of relaxation
is similar to the magnetofrictional method (\citealt{Yang1986,Ballegooijan1999})
except that the field line location, rather than local field direction,
is being relaxed. 

All relaxation codes become proportionally less stable as equilibrium
is approached, because the component forces are much larger than their
resultant, making the local linearization matrix stiff. \emph{}To
overcome this problem and prevent oscillation around the equilibrium,
FLUX scales each node's relaxation step by the square of the stiffness
coefficient $\left(\left|\sum\vec{F}_{j}\right|/\sum\left|\vec{F}_{j}\right|\right)$,
where $j$ runs over all component forces in the relaxation (in this
case the pressure and curvature forces). Further, although the forces
are field-normalized, smaller steps must be taken where the fluxons
are close together. Hence, the step law used for FLUX with field-normalized
forces is:\begin{equation}
\Delta\overrightarrow{x_{i}}=\delta\tau\left(\frac{\left|\sum\vec{F}_{j}\right|}{\sum\left|\overrightarrow{F}_{j}\right|}\right)^{2}r_{min,i}\sum\vec{F}_{j}\label{eq:step}\end{equation}
where $i$ is node number, $j$ runs over all forces included in the
relaxation, $\delta\tau$ is a small Eulerian step coefficient in
fictitious {}``relaxation time'', the central fraction is the stiffness
coefficient, and $r_{min}$ is the closest-approach distance of the
closest neighbor to the following or trailing fluxel of each node.
Relaxation continues until the stiffness coefficient falls everywhere
below some threshold, or until a maximum number of steps have been
taken.

To consider additional forces in the simulation, it is only necessary
to calculate the new force at each node at each relaxation step, and
add it to the other forces in the relaxation.

\subsection{Boundary conditions}

The end of each fluxon may be \emph{line-tied} (end node forced to
a particular location; this is the norm on the photosphere), \emph{open}
(end forced at each relaxation step to the surface of a very large
sphere), or \emph{plasmoid} (ends of the fluxon are forced to the
same location). With no further consideration for boundary conditions,
the fluxon formalism yields free-space boundary conditions, effectively
extrapolating the field to infinity \emph{in vacuo}. 

Impenetrable plane-like boundaries with prescribed field normal to
the boundary (such as the solar photosphere) are modeled by the method
of images: during each Voronoi calculation, each fluxel interacts
not only with its physical neighbors but with an image of itself reflected
through the plane of the boundary. The reflection is culled in the
Voronoi calculation process, just like any other fluxel, so that elements
far from the boundary do not interact with it directly. The image
fluxel forces edge fluxels to have a voronoi cell boundary coincident
with the boundary plane, confining the modeled flux. The field normal
to the boundary may be prescribed by tying fluxon end-points to the
boundary itself: the pattern is a constant of relaxation, and determines
the field in the vicinity of the boundary. If no fluxons are tied
to the boundary, then the field normal to the boundary is identically
zero.

Formally, there is no magnetic method of images for most curviplanar
boundaries. FLUX supports a spherical or cylindrical boundary using
a \emph{polyplanar} (or {}``disco ball'') approach, in which each
fluxel has an image that is reflected through the tangent plane directly
under the fluxel's center. This method works because the only fluxels
that interact directly with their reflections are close enough to
the surface that the boundary approximates a plane. 

\begin{figure*}[!tb]
\center{\includegraphics[%
  width=0.75\paperwidth,
  keepaspectratio]{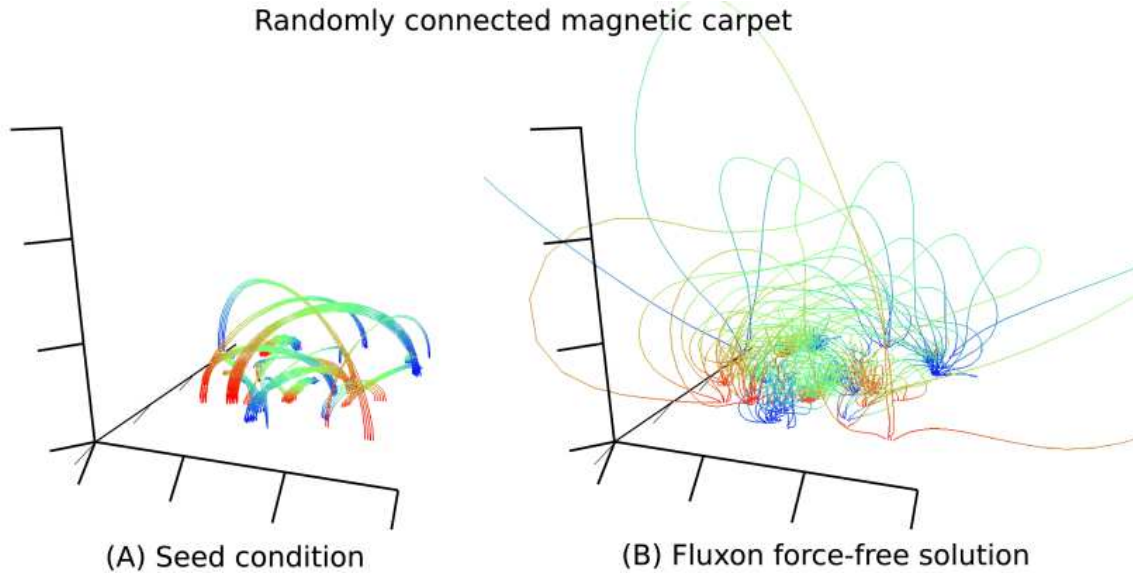}}

\caption{\label{cap:carpet}A randomly connected {}``magnetic carpet'' field
over a photosphere, calculated with FLUX\emph{.} The initial field
was seeded with a collection of flux concentrations located randomly
on a square patch of photosphere. North poles are blue; south poles
are red.}
\end{figure*}

\subsection{Initial conditions}

Because fluxon models explicitly conserve topology, the initial topology
map must be specified in advance of the relaxation, in the form of
a collection of fluxons that have the correct topology and endpoints,
but not necessarily the correct shape; this makes fluxons useful for
tracking systems with specified initial topology, such as flare models
(e.g. \citealt{Longcope1996}) or semi-empirical results from tracking
of photospheric magnetic features (e.g. \citealt{deforest2004}).
An example initial condition and resulting field solution are shown
in Figure \ref{cap:carpet}: an initial collection of fluxons is constructed
to represent the connection map and desired topology of the solution,
and then relaxed to find the actual magnetic field configuration in
the system. To solve quasi-static time dependent problems, one may
deform a relaxed solution in a nonphysical way to match updated boundary
conditions at the next time step, and then relax to find the new solution. 

Fluxon models are less directly suited to solving problems where the
initial topology is not known but the full vector field is known at
the boundary. In such cases, one must begin with a guessed initial
topology, and then use the mismatch between the computed and measured
field angle at the boundary as an error function to find the correct
topology by trial and error. Alternatively, one may use vector magnetograms
to validate topological inferences, for example those derived from
magnetic tracking and/or coronal imaging.

\subsection{Grid regularization and refinement}

Since the shape of the field lines, and not the full position of the
nodes, determines the Lorenz force, we are free to impose a nonphysical
force along the field to arrange the nodes for optimal sampling. To
ensure optimal distribution, FLUX nodes along the same fluxon
repel one another with an inverse-square law force, and also are attracted
to curvature. This results in a compromise between uniform distribution
and clustering near places where curvature is high. Every few hundred
relaxation steps, the grid is checked for denseness. Additional nodes
are placed wherever the fluxels are longer than the inter-fluxon spacing,
and wherever the turn angle at each node is too great. Currently,
there is no way to add more fluxons to a model in mid-relaxation,
though that type of grid refinement is planned for future work.

\subsection{Reconnection}

Fluxon models lock in topology, preventing any reconnection that is
not inserted explicitly into the model. Because fluxons are represented
digitially as linked lists of fluxel locations, it is simple to relink
two fluxon lists to achieve discrete reconnection in the model. FLUX
supports this capability and offers a programmer interface
to relink neighboring fluxons when particular local conditions are
met. Planned future work includes studies of current-triggered reconnection
in which, once a threshold current density is achieved, reconnection
proceeds very rapidly. This behavior is represented by the stick/slip
reconnection model of \citet{Longcope1996} and is a possible mechanism
for nanoflare heating of the coronal (e.g. \citealt{Parker1988}).

\section{\label{sec:FLUX-implementation}FLUX implementation and performance}

FLUX is written in portable C, with a user interface in Perl/PDL \citep{Glazebrook2003}.
Initial conditions may be specified either as an initial topology
map (non-equilibrium fluxon geometry) or as a collection of potential
sources together with boundary tie-points (in which case the code
shoots fluxons through the specified potential field from each tie-point).
Output is in the form of node coordinate arrays that contain the fluxon
geometry and any ancillary data (such as plasma density) indexed by
node ID number. The simulation arena is represented as a Perl object
that is manipulated using method call syntax. Individual nodes are
allocated and freed dynamically. The code can also render fluxon geometry
in 3-D using the OpenGL graphics library. Subroutines implementing
several forces (both field-normalized and non-normalized) are available
in the code, and the user can choose between them at run time. A programmer
interface exists for adding more forces into the balance. Initial
fluxon configuration and boundary conditions may be specified using
the PDL interface.

\subsection{Scaling}

While developing FLUX we noticed an interesting phenomenon in the
code's scaling properties. The number of nodes required to represent
a loop of magnetic flux is dependent on the total amount of curvature
in the loop, plus the number of inflection points in the loop. But
magnetic fields that are line-tied at the Sun's photosphere are approximately
self-similar against scaling transformations, so that large loops
require about the same number of nodes as small loops to represent.
The result is that in typical use, FLUX's memory usage depends
almost linearly on the total amount of line-tied flux that is represented,
rather than on the total simulation volume. The simulation complexity
scales more like a conventional 2-D model than like a 3-D model. We
tested this hypothesis with collections of randomly connected {}``magnetic
carpet'' fields. One such randomly generated magnetic carpet was
shown in Figure \ref{cap:carpet}, in which 40 randomly located flux
concentrations have been connected into 20 separate magnetic domains
with varying amounts of current. We generated and relaxed 20 carpet
models spanning two orders of magnitude in complexity, from 3 flux
concentrations to 300 flux concentrations, refining each so that the
final vertex angle was limited to 0.25 radian (about $15^{\circ}$)
during relaxation by 200 steps of $\Delta\tau=0.2$. We found that
the required number of nodes indeed scales as the amount of flux at
the boundary over the full range that we tested: more than two orders
of magnitude in carpet complexity.

\begin{figure}[!tbh]
\begin{center}\includegraphics[%
  width=0.90\columnwidth,
  keepaspectratio]{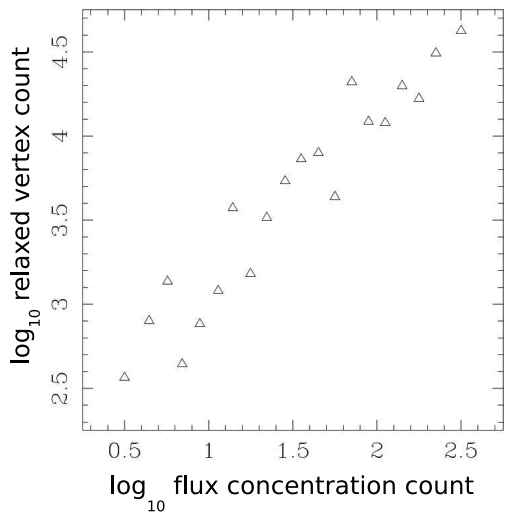}\end{center}

\caption{\label{cap:scaling}Number of nodes versus number of flux concentrations
on the boundary, for a family of magnetic carpet simulations similar
to the one in Figure \ref{cap:carpet}. Because of the field's self-similarity
against scaling, the number of nodes scales linearly with the (2-D)
complexity of the boundary although the simulations represent the
full 3-D field.}
\end{figure}

\begin{figure*}[t]
\begin{center}\includegraphics[%
  width=0.67\paperwidth,
  keepaspectratio]{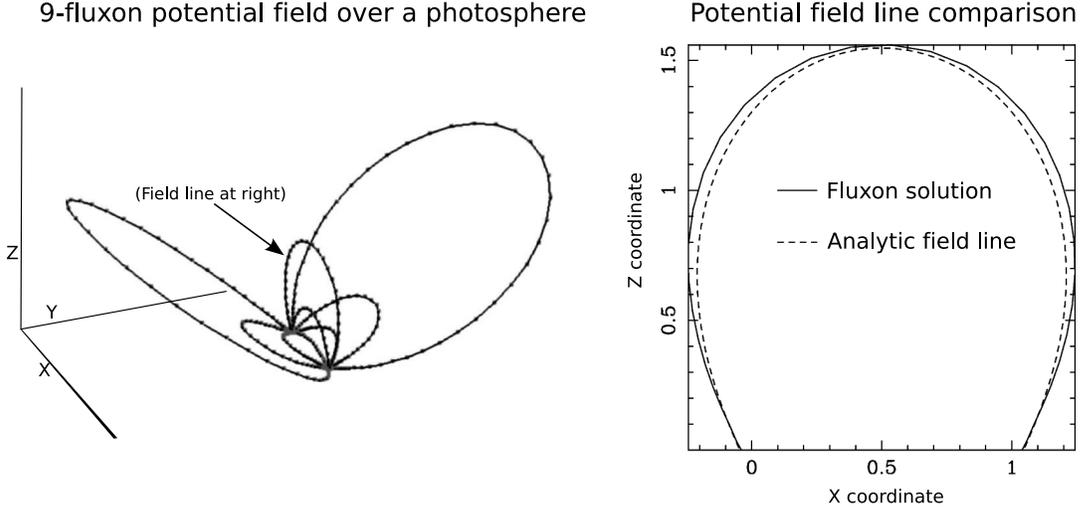}\end{center}

\caption{\label{cap:field-line-test}Simple nine-fluxon relaxation reproduces
a potential field. \textbf{LEFT:} 3-D rendering of the final fluxon
configuration. \textbf{RIGHT}: the central vertical field line closely
approximates the shape determined by shooting through an analytic
field, as seen in this rendering of the XZ plane. All nine fluxons
achieve similar matches in shape.}
\end{figure*}

\subsection{Simple validation cases}

We demonstrate here that FLUX can reproduce a simple potential field
calculation and the behavior of the \citet{Lundquist1950} (linear
force-free) and \citet{GoldHoyle1960} (nonlinear force-free) flux
tubes.

\paragraph{Potential solution}

Because FLUX is a full nonlinear force-free field solver, the
potential solution has no special properties for testing the code,
aside from convenience and ease of representation. We used nine fluxons
to connect a square grid of tie points with spacing 0.05, centered
at location (0,0,0), to a similar grid at (1,0,0), with a photosphere
located in the XY plane. Free-space boundary conditions apply elsewhere
than the photosphere, allowing the solution to expand into the positive-Z
half-space. Each fluxon's initial configuration was a single upward
jump followed by five horizontal steps and a downward jump (8 nodes
per fluxon, including endpoints). The fluxons were relaxed for 1,000
timesteps with $\delta\tau$ set to 0.2, with nodes added periodically
to limit the inter-fluxel angle to 0.1 radian (about $6^{\circ}$).
The relaxation required about 40 CPU-seconds on a 1.4 GHz Athlon workstation,
ending with just under 700 nodes total and an average stiffness coefficient
of $8\times10^{-3}$, indicating good relaxation. The resulting potential
field approximation is rendered in Figure \ref{cap:field-line-test}.
The tied line locations approximate the surface penetration of a finite
dipole with poles at (0,0,-0.1) and (1,0,-0.1), so we compared the
final relaxed shape of each fluxon to the shape of an analytically
calculated field line with the same footpoints. The match is quite
good, especially considering the coarseness of the model: $\left|\Delta\overrightarrow{B}\right|/B$
between the analytic and fluxon solution is under 8\% at every fluxon
node, and under 2\% everywhere except within 0.05 of the footpoints.

\paragraph{Linear force-free (Lundquist) flux tube}

To demonstrate that FLUX is capable of matching analytic force-free
solutions that are not potential, we demonstrate convergence to the
\citet{Lundquist1950} solution in linear geometry. The Lundquist
solution is a cylindrical twisted flux tube along the $z$ axis, with
the form \begin{equation}
B_{z}=B_{0}J_{0}(kr);\, B_{\theta}=B_{0}J_{1}(kr);\, B_{r}=0.\label{eq:Lundquist}\end{equation}
We seeded the solution by shooting fluxons through the analytic Lundquist
solution with $k=1$, between the $z=0$ and $z=30$ planes, for $r$
values from $0$ to 10 (near the third node of $J_{1}$). At each
$dr$ step in radius, we computed the number of field lines penetrating
the $z=0$ plane at that diameter and truncated to the nearest integer,
saving the floating-point residual to add at the next $dr$. The integer
number of field lines were launched equidistant in $\theta$ around
the flux tube, with a phase shift of 2.0 radians at each $dr$. This
field line placement has the correct field topology by construction,
but is far from correct: chance groupings of field lines in the essentially
randomly oriented cylindrical shells dominate variations in field
strength.

The resulting paths were inserted into a fluxon model that was relaxed
for 300 steps with $\delta\tau$ set to 0.2, with nodes added periodically
to limit the inter-fluxel angle to 0.2 radian. Because $\int xJ_{n}^{2}dx$
diverges, the Lundquist flux tube cannot be reproduced with free-field
boundary conditions and a finite amount of flux; hence, we applied
a low-beta cylindrical boundary at $r=10.05$ during the relaxation:
among the neighbor candidates for each fluxel at each time step was
an {}``image fluxel'' created by reflecting each of the fluxel's
two vertices through the plane tangent to the cylinder radially outward
from the corresponding vertex. This prevented flux tube expansion
through the free space around the cylinder, while not directly affecting
any flux element in the interior of the modeled tube.

\begin{figure*}[!t]
\center{\includegraphics[%
  width=0.75\paperwidth,
  keepaspectratio]{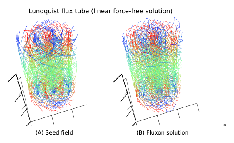}}

\caption{\label{cap:Lundquist}Three-dimensional rendering of the Lundquist
flux tube solution with FLUX. Analytically computed field lines
were generated and converted to fluxons (LEFT), then perturbed randomly
and relaxed with FLUX (RIGHT) in the presence of a cylindrical
impermeable boundary. Field line direction is rendered in color, grading
from blue at north magnetic poles to red at south magnetic poles.}
\end{figure*}
\begin{figure*}[!t]
\center{\includegraphics[%
  width=0.67\paperwidth,
  keepaspectratio]{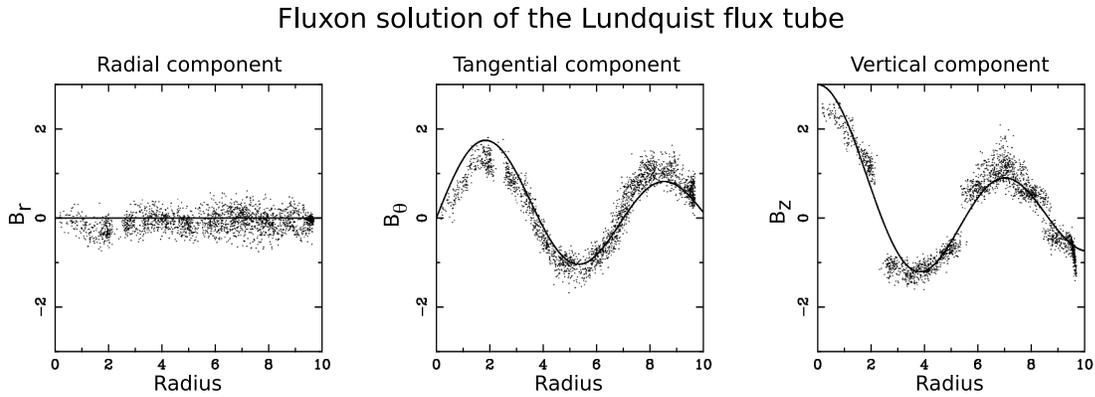}}

\caption{\label{cap:Lundquist-plots}B-field components in cylindrical coordinates
at each fluxel center location in the Lundquist flux tube solution
in Figure \ref{cap:Lundquist}. The theoretical values are plotted
as solid lines. }
\end{figure*}

The relaxation involved 200 fluxons anchored in 50 concentric rings,
consumed about 1000 CPU-seconds, and ended with just over 7,000 nodes
and an average stiffness coefficient of $2\times10^{-2}$ . The resulting
approximation of the Lundquist flux tube is rendered in Figure \ref{cap:Lundquist}.
In cylindrical coordinates, the Lundquist solution is dependent only
on $r$, the distance from the $z$ axis; hence, to determine the
quality of the solution, we produced scatterplots of the calculated
field components at each fluxel center in the relaxed solution. Only
fluxels in the central third of the solution ($10\leq z\leq20)$ were
considered, to reduce edge effects at the top and bottom of the fluxon
system. The scatterplots are shown in Figure \ref{cap:Lundquist-plots},
demonstrating that the solution has converged to the Lundquist solution.
The RMS $\left|\Delta\vec{B}\right|/B$ is 5.4\% throughout the volume.

Two features of the plots in Figure \ref{cap:Lundquist-plots} require
some explanation. First, there is considerable scatter in the plots
compared to the potential solution. The scatter is attributable to
two separate effects. First, because of the small number of fluxons
in each radial sheath, there is some distortion of the field as the
fluxons bend one around the other. Secondly, we are representing the
field within each fluxon using sectorwise angular equipartition --
essentially assuming that the currents are concentrated along the
boundaries of triangular prisms around each fluxon -- but the Lundquist
solution requires a volume current. Increasing the number of fluxons,
or using an interpolation technique that averages the field over more
than one fluxel, reduces the scatter significantly both by reducing
the bending effect and by better approximating a volume current. 

Secondly, the $z$ component of the field strength jumps across zero
near the nodes of $J_{0},$ rather than passing smoothly through it.
This is due to the discrete nature of the fluxons and the construction
of our initial seed solution: no fluxons were launched from the rings
where $J_{0}=0$, because by construction each ring only contained
sufficient fluxons to approximate the total magnetic flux penetrating
the ring. Hence our solution develops a small current sheet near each
node of $J_{0}$. While better attention to the seed field would reduce
this artifact, we have retained it here to point out the care that
is required in selecting the initial seed field and fluxon locations.
The gaps in the statistical population between $r=2.5$ and $r=5.5$
are due in part to this effect and in part to the periodicity in $r$
of the original seed population of fluxels; this fossil periodicity
is more readily apparent in the radial component scatterplot at far
left.

While we have plotted field values only at fluxel centers, the field
can be calculated anywhere within the simulation volume by interpolating
between the field values at nearby fluxels.

\paragraph{Nonlinear force-free (Gold-Hoyle) flux tube}

The Gold-Hoyle flux tube has the interesting property that $d\theta/dz$
is constant across field lines. The Gold-Hoyle solution has the form\begin{equation}
B_{z}=\frac{B_{0}}{1+\mu^{2}r^{2}};\, B_{\theta}=\frac{B_{0}\mu r}{1+\mu^{2}r^{2}};\, B_{r}=0.\label{eq:gold-hoyle}\end{equation}
 which, like the Lundquist solution, carries an infinite amount of
flux (the enclosed flux diverges logarithmically in $r$). We tested
a Gold-Hoyle-like solution by launching a square, 11x11 array of fluxons
from each of two flux concentrations in free space, separated by a
distance of 30 on the $z$ axis. The flux concentrations were parallel
to the $xy$ plane, and had an inter-fluxon spacing of just 0.1. The
total twist was 2.0 turns along the length of the flux concentration.
Because the twist is locked in by the fixed topology, this nonphysical
initial condition should relax to become a Gold-Hoyle flux tube with
the correct amount of twist for its final radius. We allowed it to
expand to a radius of 7.5 using zero-normal-field cylindrical boundary
conditions. Figure \ref{cap:gold-hoyle-plots} shows the seed field
configuration and the relaxed flux tube. We fit compared the fluxon
field results to the family of analytic solution by plotting field
component values at each fluxel center in the middle third of the
simulation volume, and fit the value of $\mu$ by eye to about 0.65.
The results are shown in Figure \ref{cap:gold-hoyle-plots}. 

As with the Lundquist solution, we have plotted the magnetic field
value only at fluxel centers. Figure \ref{cap:gold-hoyle-plots} shows
a good match between the fluxon result and the analytic solution,
with no large current sheets as were demonstrated in the Lundquist
flux tube. As above, the periodicity of the fluxon placement is due
to the regularity of the initial seed field, but the field strength
and direction are close to the analytic value everywhere. The fossil
periodicity is much more evident in this solution than in the Lundquist
solution, because the regularity of the initial condition ensures
less relative movement of the fluxons in finding their equilibrium
positions. 

The slope of $B_{\theta}$ vs. $r$ is slightly steeper in the fluxon
curve than in the analytic solution. We attribute this effect to the
vertical expansion of the outermost fluxons outside the intended cylindrical
volume, which reduces the amount of twist per unit length in the outer
portion of the volume. The effect may be reduced by expanding the
source pattern, as in the seed condition for the Lundquist solution,
above, or by imposing impenetrable boundary planes above and below
the tube.

\begin{figure*}[!tb]
\center{\includegraphics[%
  width=0.75\paperwidth,
  keepaspectratio]{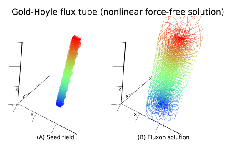}}

\caption{\label{cap:gold-hoyle}Three-dimensional rendering of the Gold-Hoyle
flux tube solution with FLUX. Analytic nonphysical field lines were
generated and converted to fluxons (LEFT), then relaxed with FLUX
(right) in the presence of a cylindrical impermeable boundary. Field
line direction is rendered in color, grading from blue at north magnetic
poles to red at south magnetic poles.}
\end{figure*}

\begin{figure*}[tb]
\center{\includegraphics[%
  width=0.67\paperwidth,
  keepaspectratio]{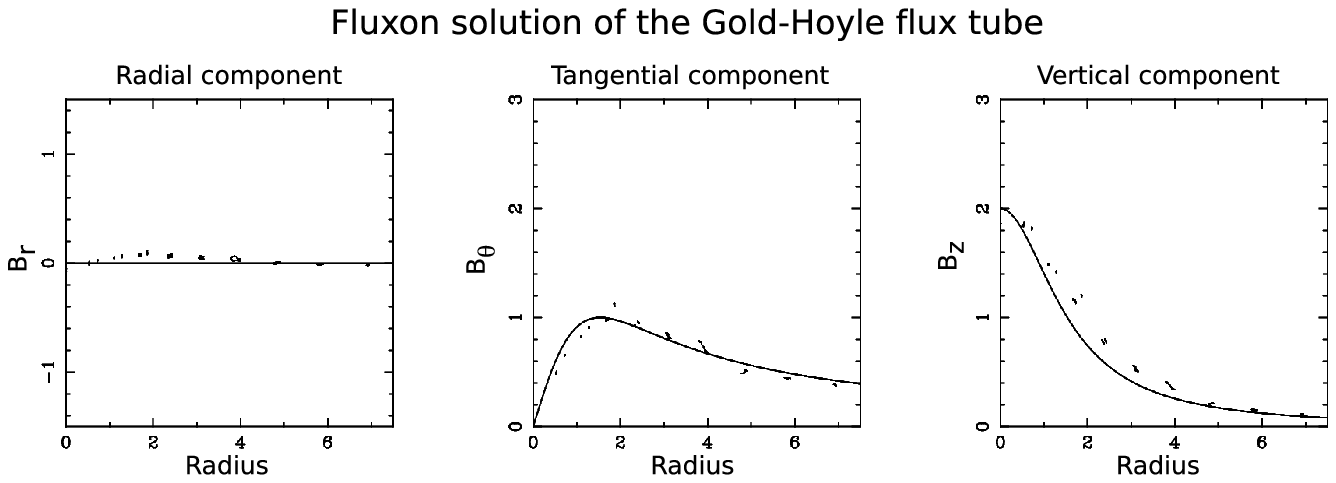}}

\caption{\label{cap:gold-hoyle-plots}B-field components in cylindrical coordinates
at each fluxel center location in the Gold-Hoyle flux tube solution
in Figure \ref{cap:gold-hoyle}. The theoretical values are plotted
as solid lines.}
\end{figure*}

\section{\label{sec:Future-work}Conclusions and future work}

We have introduced \emph{fluxon modeling} and a prototype code, FLUX,
that is being released as free software; and have demonstrated that
FLUX can reproduce simple potential, linear, and nonlinear force free
field solutions. FLUX is currently useful as a magnetofrictional force-free
field solver, but it is also intended as a prototype of a much more
complete MHD model. FLUX is very promising in two important respects:
first, it exactly preserves field topology, potentially yielding a
better approximation of ideal MHD than is possible with a conventional
Eulerian approach to MHD; and second, it demonstrates good scaling
properties that suggest it will perform very well when applied to
more complex systems.

We are releasing FLUX as free software that is available as source
code for any purpose at all. It may be obtained from the authors,
via the web at \texttt{\small http://www.boulder.swri.edu/\textasciitilde{}deforest/FLUX},
via Solarsoft \citealt{Freeland1998} as part of the {}``PDL'' package,
or via the Community Coordinated Modeling Center \citealt{Hesse2002}.
Future work on FLUX will take two directions: addition and testing
of plasma and other forces to study non-force-free equilbria; and
addition of dynamic forces to study quasi-stationary MHD systems and,
ultimately, full inertial MHD evolution.

\acknowledgements{Thanks go to many people without whom FLUX could not have been developed.
D. Longcope of Montana State University provided inspiration and connected
the authors, each of whom had begun work independently; T. Bogdan
of the High Altitude Observatory provided good insight and discussion
about line-tied boundary conditions and stability criteria; T. Munzner
introduced us to the concept of Voronoi analysis. Thanks, also, to
S.T. Wu and J. Spann, the editors, for kindly allowing us a few extra
pages, and to the referees both for their insightful comments and
for pointing out (correctly) that we needed more length. FLUX was
developed under grant from NASA's LWS TR\&T program. Continued fluxon
model development is being funded internally by Southwest Research
Institute and by NASA's LWS TR\&T program.}

\bibliographystyle{plainnat}

\end{document}